\documentclass[journal]{vgtc}                

\ifpdf
  \pdfoutput=1\relax                   
  \usepackage{graphicx}                
  \DeclareGraphicsExtensions{.pdf,.png,.jpg,.jpeg} 
\else
  \usepackage{graphicx}                
  \DeclareGraphicsExtensions{.eps}     
\fi%

\graphicspath{{figures/}{pictures/}{images/}{./}} 

\usepackage{microtype}                 
\PassOptionsToPackage{warn}{textcomp}  
\usepackage{textcomp}                  
\usepackage{mathptmx}                  
\usepackage{times}                     
\usepackage{cite}                      
\usepackage{tabu}                      
\usepackage{booktabs}                  

\usepackage{multirow}
\usepackage{verbatim} 
\usepackage{enumitem}
\usepackage{amsmath,amssymb}
\usepackage{hyperref}

\usepackage{flushend}

\onlineid{0}

\vgtccategory{Research}
\vgtcpapertype{please specify}

\title{Polyphorm: Structural Analysis of Cosmological Datasets \\via Interactive Physarum Polycephalum Visualization}

\author{Oskar Elek, Joseph N. Burchett, J. Xavier Prochaska, and Angus G. Forbes}
\authorfooter{
\item
 O. Elek and A. G. Forbes are with the Dept. of Computational Media at University of California, Santa Cruz. Email: \{oelek,angus\}@ucsc.edu.
\item
 J. N. Burchett and J. X. Prochaska are with the Dept. of Astronomy and Astrophysics at University of California, Santa Cruz. Email: \{burchett,xavier\}@ucolick.org.
}

\shortauthortitle{Elek \MakeLowercase{\textit{et al.}}: Polyphorm}

\abstract{This paper introduces \textit{Polyphorm}, an interactive visualization and model fitting tool that provides a novel approach for investigating cosmological datasets. Through a fast computational simulation method inspired by the behavior of Physarum polycephalum, an unicellular slime mold organism that efficiently forages for nutrients, astrophysicists are able to extrapolate from sparse datasets, such as galaxy maps archived in the Sloan Digital Sky Survey, and then use these extrapolations to inform analyses of a wide range of other data, such as spectroscopic observations captured by the Hubble Space Telescope. Researchers can interactively update the simulation by adjusting model parameters, and then investigate the resulting visual output to form hypotheses about the data. We describe details of \textit{Polyphorm}'s simulation model and its interaction and visualization modalities, and we evaluate \textit{Polyphorm} through three scientific use cases that demonstrate the effectiveness of our approach.%
} 


\CCScatlist{ 
 \CCScat{K.6.1}{Management of Computing and Information Systems}%
{Project and People Management}{Life Cycle};
 \CCScat{K.7.m}{The Computing Profession}{Miscellaneous}{Ethics}
}

\teaser{
  \centering
  \includegraphics[width=\linewidth]{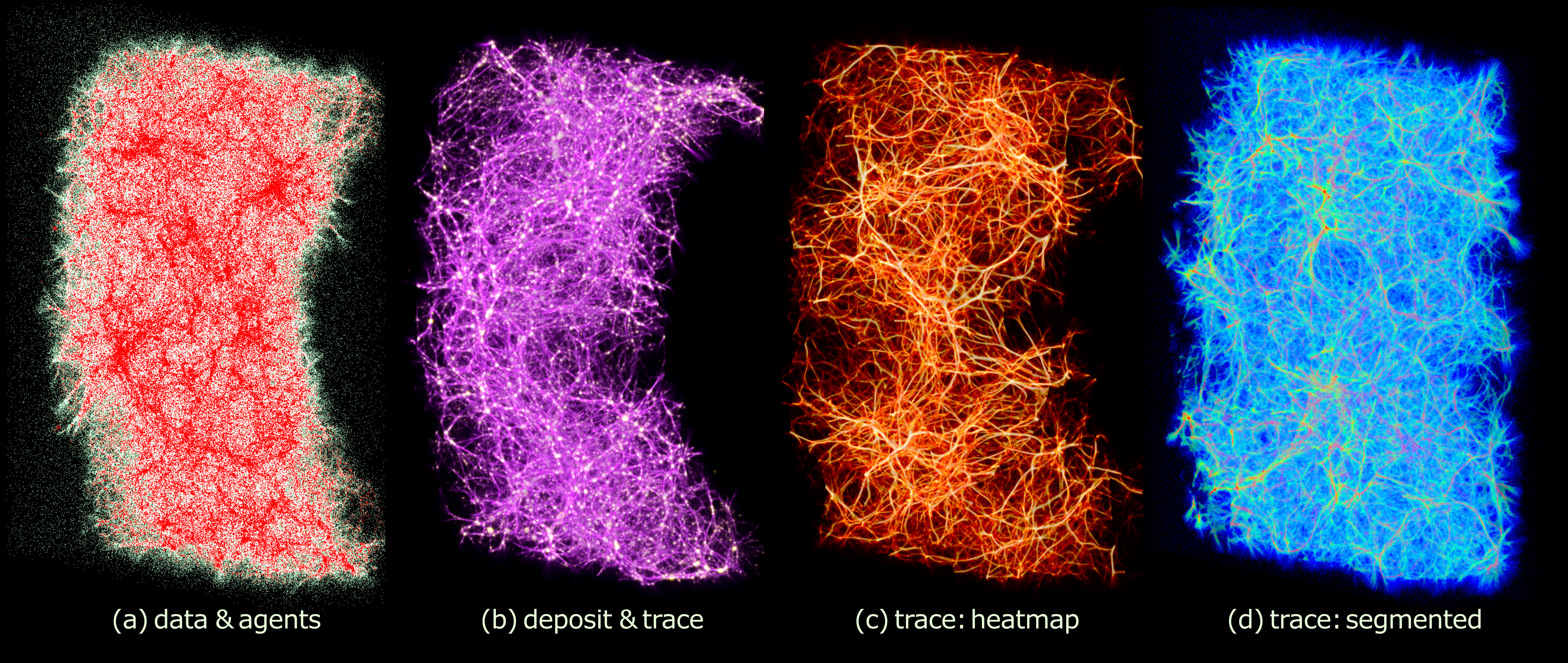}
  \vspace{-6mm}
  \caption{This figure depicts the use of the \textit{Polyphorm} software tool for interactive visualization and data reconstruction during an analysis of the Bolshoi-Planck cosmological simulation dataset, which consists of approximately 840,000 simulated dark matter halos spanning a region of 170 Mpc in each dimension. Here, the entire volume is split into 4 equally sized vertical slabs, each showing a different view that enables the analysis of dark matter filaments: (a) raw data points (red) and agents (white) of our Monte Carlo Physarum Machine (MCPM) reconstruction algorithm, (b) volumetric footprint of the data (`deposit’, white) and the reconstructed Cosmic Web filament density field (`trace’, purple), (c) density field of the reconstruction mapped using a heatmap color palette, and (d) the same density field, but now segmented instead into three intervals using an alternative color map (`low’ in blue, `medium’ in green, `high’ in red). The filament reconstruction and interactive visualization facilitates new insights into into the structure and composition of the Cosmic Web. }
  \label{fig:teaser}
}

\begin{document}


\firstsection{Introduction}

\maketitle

Composed primarily of hot gas and plasma, the intergalactic medium (IGM) is theorized to contain a significant amount of the matter in the Universe that has so far eluded observation. Astrophysicists refer to this as ``the missing baryon problem''~\cite{Fukugita1998}, and the detection and localization of IGM remains an ongoing research effort. The best evidence that the IGM contains the unaccounted baryonic matter is provided by large-scale cosmological simulations, such as the Millenium~\cite{Springel2005}, EAGLE~\cite{schaye2015eagle} and Bolshoi-Planck~\cite{Klypin2016} simulations. These simulations also unanimously agree on the expected distribution of the IGM: after billions of years, the combined effect of gravitational attraction and dark-energetic repulsion has shaped the universe as a vast network called the Cosmic Web. This quasi-fractal structure~\cite{Scrimgeour2012} consists of knots--- galaxies and their clusters--- interconnected by a complex network of filaments and separated by vast cells of empty space called voids. In addition, the entire structure is underlined by a scaffolding of dark matter. These simulations have been constructed to confirm the earlier findings of the Zeldovich framework~\cite{Zeldovich1970, Doroshkevich1970, Icke1973}, which on theoretical grounds predicts the emergence of these very structures--- knots, filaments, voids--- under the influence of the dominant force dynamics in the Universe.

Given the indirect evidence for its existence, astrophysicists are interested in determining the structure of the Cosmic Web. The intergalactic filaments consist primarily of diffuse non-luminous matter, and are extremely challenging to observe directly. Both topological~\cite{Aragon-Calvo2010} and geometric~\cite{Tempel2014, Chen2016} approaches using graphs as a core representation to model the Cosmic Web have been proposed. However, most of the existing approaches focus on cataloging and classifying the filaments~\cite{Libeskind2018}, and no method described to date has been able to sufficiently recover the full 3D density map of the intergalactic filaments. Given that the Cosmic Web is a heterogeneous multi-scale structure without an intrinsic topology, having an accurate estimate for its density distribution would benefit many tasks, such as a better understanding the transition from the galactic to intergalactic environments~\cite{Burchett2020}, and shed light on the missing baryon problem itself.

The Physarum polycephalum organism~\cite{guttes1961morphological}, a type of slime mold, has been used as an unconventional ``biological computer'' to solve a range of spatial problems~\cite{Adamatzky2010, Adamatzky2016}, leveraging the organism’s propensity to systematically explore its environment for food and to grow intricate yet natural networks~\cite{Whiting2016, Sun2017}. Examples include maze solving~\cite{Nakagaki2000}, shortest path finding~\cite{Nakagaki2001}, transportation system design~\cite{Tero2010} and the travelling salesman problem~\cite{Jones2014a}, among others. Computational models of Physarum polycephalum, or virtual Physarum machines, have also been developed to model a range of problems, mimicking the organism's foraging behavior, and adapting it to new contexts. Food sources thus become straightforward proxies for input data, and different chemical and physical stimuli are available as tunable parameters that steer the Physarum's growth. 

In this paper, we introduce Monte Carlo Physarum Machine (MCPM), an agent-based virtual Physarum machine that algorithmically mimics the organism's growth and is able to construct precise trajectories, and the \textit{Polyphorm} interactive visualization software, which facilitates the interactive analysis of large cosmological datasets using MCPM. Relying on Physarum's ability to approximate optimal transport networks~\cite{zhang2015biologically}, we ``feed'' to its virtual counterpart galaxies and dark matter halos as 3D point attractants. The MCPM agents navigate these input points to find most probable paths between them, and the total aggregate of their trajectories then forms the filamentary network. We capture these trajectories as a density field in 3D space, interpreting it as a proxy of the intergalactic gas distribution that provides an estimate for the Cosmic Web. Using \textit{Polyphorm}, astrophysicists are able to gain new insights into the composition and structure of the Universe. 

\begin{figure}[tb]
 \centering
 \includegraphics[width=0.9\columnwidth]{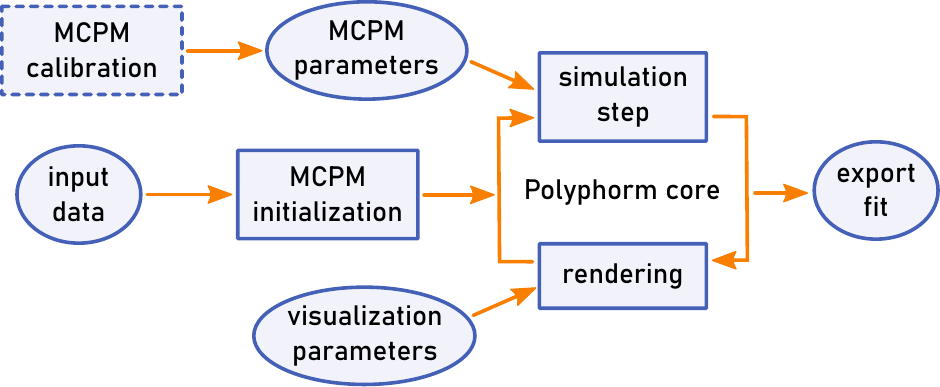}
 \vspace{-3mm}
 \caption{Overview of \textit{Polyphorm}'s pipeline. The core of the software relies on a tight coupling between the simulation and visualization components, both of which can be interactively controlled to enable expert supervision of the reconstruction process.}
 \label{fig:pipeline}
 \vspace{-5mm}
\end{figure}

The contributions of this paper are: a) an exposition of the MCPM model, b) an introduction to the \textit{Polyphorm} software and a discussion of the visualization techniques that enable the application of MCPM to cosmological datasets and the analysis of the simulation outputs, and c) a series of scientific use cases that demonstrate the utility of \textit{Polyphorm} in predicting the location and density of IGM filaments. In Sec.~2, we contextualize this work in terms of previous related work in both the visualization community and the astrophysics community. In Sec.~3, we present a taxonomy of the primary structural and visual analysis tasks enabled through our approach, as well as the design requirements for our software implementation. In Sec.~4, we present details of the MCPM, focusing on innovation in the agent propagation step, which is defined in terms of different probability distributions. In Sec.~5, we provide a summary of the visual analysis features available in the \textit{Polyphorm} software. In Sec.~6, we evaluate our methods by demonstrating how the use of \textit{Polyphorm} facilitates the investigation of the filamentary structure of the Cosmic Web. A schematic providing a high-level overview of \textit{Polyphorm} is shown in Fig.~\ref{fig:pipeline}. 

\section{Related Works}

Cosmological simulations based on the cold dark matter paradigm predict that matter in the Universe is organized into a Cosmic Web, a series of interconnected filaments that are formed from dark matter comprised of low density IGM gas as well as galaxies and the circumgalactic medium (CGM) gas found within dark matter halos adjacent to these galaxies~\cite{Cen:1994aa,Bond:1996aa,Rauch:1998aa,tumlinson2017circumgalactic}. In regions of the Universe nearer to the Milky Way, the distribution of galaxies mapped by large surveys visually indicate the Cosmic Web structure~\cite{Geller:1989aa}, and a study by Wakker et al.~\cite{wakker2015nearby} uses measurements from the Hubble Space Telescope to investigate one visually identified Cosmic Web filament and its imprint of H I absorption lines.

These cosmological simulations attempt to model physical processes spanning several orders of magnitude in scale and use empirical observations to validate the models' accuracy. While the models are tuned to match certain aspects of the observed galaxy population, other observable quantities may serve as pure predictions (consequences of the physics being tuned to match certain data). Thus, as new measurements become available, astrophysicists can refine the simulations, leading to a better understanding of the physical processes in galaxy formation. For example, the EAGLE project~\cite{schaye2015eagle} presents hydrodynamical simulations that model the formation and co-evolution of galaxies and black holes with the largest scale structures of the Universe. The FIRE project~\cite{hopkins2018fire} also produces simulations of galaxy formation, but focuses on resolving the interstellar medium of individual galaxies. The Hardware/Hybrid Accelerated Cosmology Code (HACC) framework~\cite{habib2016hacc} includes models of baryonic matter as well as active galactic nuclei associated with violent bursts of energy from supermassive black holes, and has been used to illustrate how this energy is imparted to the surrounding gas and affects subsequent structure formation~\cite{hesse2019interactive, nguyen2019visualization, fritschi2019visualizing, schatz2019visual}.

Kapferer and Riser~\cite{kapferer2008visualization} articulate the importance of interactive visualization to help researchers understand the coherences in cosmological simulations, and a range of frameworks have been developed to support astrophysics visualization. For example, Woodring et al.~\cite{woodring2011analyzing} introduce new analysis features implemented within ParaView~\cite{ahrens2005paraview,ayachit2015paraview} to analyze simulation data, and the Aladin Sky Atlas~\cite{bonnarel2000aladin} enables users to add annotation markers to observational image data catalogs. More recently,  the Open Space ``astrographics'' system~\cite{bock2019openspace} facilitates the interactive display of integrated data from multiple sources, supporting research and science communication. A number of tools present celestial objects as volumes within a 3D view~\cite{popov2012analyzing,punzo2015role,fu2007transparently,taylor2017visualizing}, and Pomar{\`e}de et al.~\cite{pomarede2017cosmography} make use of images, videos, and derived isosurface structures within a 3D representation to show galaxy position, density fields, gravitational potential, and velocity shear tensors. Visual analytics software has been developed to explore simulation data that models the evolution of the Universe~\cite{hanula2015cavern,preston2016integrated}. For example, Almryde and Forbes~\cite{almryde2015halos} introduce an interactive web application to visualize ``traces'' of dark matter halos, and an application developed by Scherzinger et al.~\cite{scherzinger2017interactive} provides 2D and 3D views to support the analysis of halo substructures and hierarchies.

To the best of our knowledge, no previous techniques explore cosmological data using a virtual Physarum machine. However, there are a range of works that utilize models of Physarum polycephalum to make sense of complex systems. Efficient virtual Physarum machines have been developed that utilize cellular automata~\cite{Kalogeiton2015, Dourvas2019} and formal logic~\cite{Schumann2016}, and an approach introduced by Jones~\cite{Jones2010} uses a large number of agents to generate emergent structures. This latter approach is similar in some ways to agent-based models that model crowds and flocks of birds~\cite{Reynolds1987, Olfati-Saber2006}, but differs in that it uses a continuous representation of the agents’ density for navigation~\cite{Jones2010, Jones2015}. In addition to the successful application to ``classic'' problems such as road network optimization~\cite{Adamatzky2017}, the agent-based computational approach is suitable for abstract and interactive tasks, such as B-spline fitting~\cite{Jones2014}, convex and concave hull approximation~\cite{Jones2014} and robot navigation~\cite{jones2015multi}. The design of \textit{Polyphorm} is also inspired by an expressive model--- itself based on Jones' pattern formation explorations~\cite{Jones2010}--- created by the artist Sage Jenson as a means for producing intriguing, complex animations~\cite{JensonWebsite}.

The structure-finding approach of \textit{Polyphorm} provides a coupling between the reconstruction of the 3D density field and its visualization, enabling researchers to interact with the model both visually and through numerical measures simultaneously. Since the underlying model (see Sec.~\ref{Sec:SimulationAndModelFitting}) runs interactively on a desktop computer with a dedicated GPU, the fitting process takes at most a few minutes, allowing for interventions, whereas previous filament finding methods~\cite{Tempel2014, Libeskind2018} can take significantly longer and require access to high performance computing systems. Contrary to approaches that produce either graph-based reconstructions or classification masks~\cite{Libeskind2018,Aragon-Calvo2010}, \textit{Polyphorm} constructs a continuous network-like geometric structure in which densities are defined at every point in the 3D volume, and where the derived density field may be validated visually against the input data distribution and weights. The availability of such density data greatly expands the range of questions we can now ask about the intergalactic medium~\cite{Burchett2020,burchett2019igm}.

\section{Task Taxonomy}
\label{Sec:TaskTaxonomy}
Based on our review of the literature and on extensive conversations with astrophysicists researching intergalactic environments and interested in understanding the structure and composition of the Cosmic Web, we define the main structural analysis tasks that our model supports. Additionally, to facilitate structural analysis, we further define a set of visual analysis tasks that support the investigation and interpretation of the output generated via MCPM.

\subsection{Structural Analysis Tasks}

\noindent\textbf{S1: Generate density fields.} A core task for astrophysicists investigating dark matter filaments and the intergalactic medium that permeates them is to gain insight into the structure of the Cosmic Web. The primary output of \textit{Polyphorm} is a 3D density field that corresponds to (or ``reconstructs'') the input data---a set of points in 3D Euclidean space, each optionally assigned a weight---in a meaningful way. In the astronomical context, that conforms to the rules of optimal transport. While there are many approaches to finding higher-order structures in point data, many of them are based on discrete, graph-like representations~\cite{benson2016higher}. Though these representations enable easy navigation and organization, they can obscure important aspects of some astrophysics datasets.

In the scientific use cases described in Sec.~\ref{Sec:UseCases}, a key dataset is the Bolshoi-Planck simulation, which explores dark matter dynamics in a cubic region of roughly 200 Mpc in size. (For reference, the Milky Way is 0.03 Mpc in diameter). Out of this massive dataset, a set of about 2M dark matter halos has been extracted and are used as the basis for our fitting. The presence of ground-truth densities enables the calibration of our model in terms of mapping to the canonical ``cosmic overdensity'' quantity. The other key dataset is the catalog of observed galaxies mapped by the Sloan Digital Sky Survey (SDSS). Querying observations from the range of cosmic redshifts $z = 0.0138-0.0318$ yields approximately 37k galaxies, again suitably represented by 3D weighted points across the 200 Mpc region of interest. In contrast to the simulated data, the observations are known to be incomplete, and lack a ground-truth density map. By generating reasonable density fields, astrophysicists can better understand the structure of the Cosmic Web.

\ \\ \noindent\textbf{S2: Identify anisotropic structures.} Once an estimated density field is available, astrophysicists can use it to identify anisotropic structures of the Cosmic Web. Such structures are expected to correspond to the dark matter filaments and should in turn comprise the gaseous filaments of the intergalactic medium. The presence of these structures is predicted by simulations~\cite{Springel2005,schaye2015eagle,Klypin2016}, resulting mainly from the gravitational collapse of matter on the Universal timescales. Unfortunately, observational evidence of these structures is extremely challenging to obtain due to the low particle density of these filamentary structures. One notable exception are spectroscopic measurements parametrized by redshift (distance from Earth), which systematically show absorption signatures even in regions of space where galactic halos and other visible agglomerations of matter are not present. The key ingredient in finding these structures is a model capable of consistently reconstructing such structures from sparse positional data.

\ \\ \noindent\textbf{S3: Detect features across different scales.} A task related to \textbf{S2} is detecting geometric features across several orders of magnitude, in terms of both spatial scale and density. Estimations of the scale of the Cosmic Web filaments range between 1 and 100 Mpc. Smaller filaments contain less matter, so the reconstructing model needs to generate a density field with several orders in dynamic range. This enables researchers to simulate both dense and sparse regions of the Universe.

\ \\ \noindent\textbf{S4: Support structural transfer between relevant datasets.} Another important task is supporting structural transfer between input point clouds. The idea is to use structures detected in one dataset as a prior estimator for the model parameters when fitting to another dataset. This task becomes useful with the availability of simulated data, as structures detected in a dense simulated dataset (especially with ground-truth density available) can bootstrap the fitting to observation data. Doing so is expected to impact the robustness of the model when fitting to potentially incomplete data, as well as the confidence in such a fit.

\begin{figure*}[tb]
 \centering
 \includegraphics[width=0.92\textwidth]{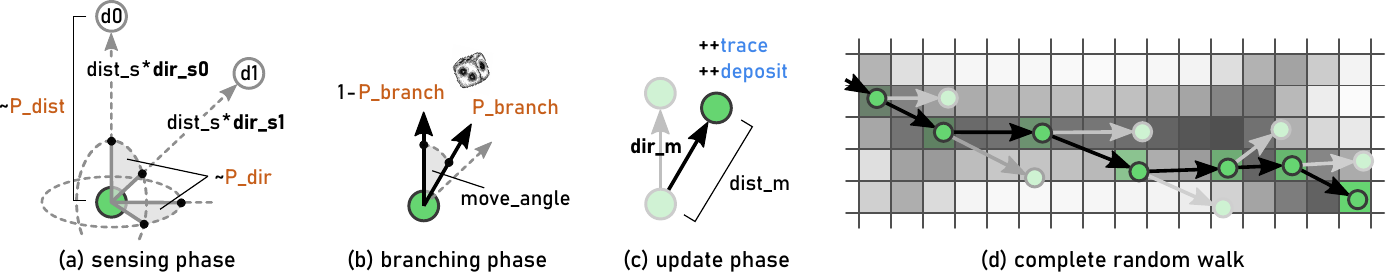}
 \vspace{-2mm}
 \caption{Visual depiction of the low-level sampling decisions performed by each MCPM agent per single propagation step. All three steps (a)--(c) are stochastic, generating a random walk (d) which constitutes the agent's trajectory guided by the data-emitted deposit (grey field).}
 \label{fig:prop_steps}
 \vspace{-3mm}
\end{figure*}

\begin{figure}[b]
 \centering
 \vspace{-3mm}
 \includegraphics[width=\columnwidth]{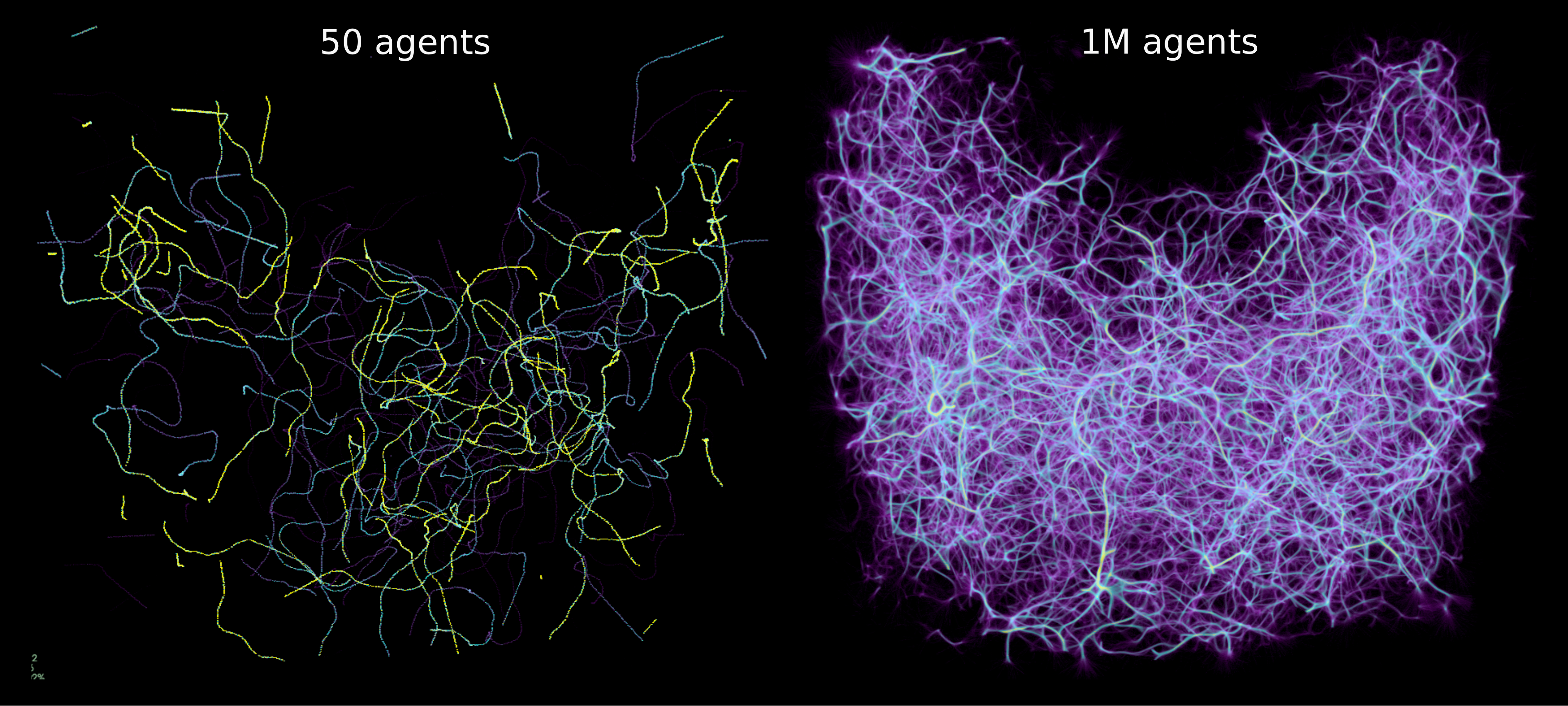}
 \vspace{-6mm}
 \caption{Superposition of agent trajectories in 3D space. While individual agents follow seemingly chaotic paths, the order emerges from a statistical aggregate of millions of them. That aggregate--- the \textit{trace} field--- is the reconstructed output of MCPM.}
 \label{fig:agent_trajectories}
\end{figure}

\subsection{Visual Analysis Tasks}
\label{Sec:VisualAnalysisTasks}
Complementing the structural analysis tasks, we identify a number of important tasks that support the visual analysis of the model. These tasks facilitate the understanding of the input point data distribution, as well as provide diagnostics of the Cosmic Web estimate, including its global and local structural properties, alignment with the input data, and allocation of matter density. Visual feedback is especially important here, since the solution to the Cosmic Web reconstruction problem is heavily under-determined. That is, we cannot rely only on numerical fitting, since the loss function that would define the fitness of the solution is not initially known.

\ \\ \noindent\textbf{V1: Interactive exploration of the input data and MCPM simulation.} The core task here is to visualize both the input data and the resulting Cosmic Web estimate. Interacting with the 3D rendering is necessary due to the supervised nature of the reconstruction methodology, and researchers need to be able to freely rotate, pan, and zoom into the data and the solution. An essential part of this task is the ability to slice the 3D data, given the presence of many overlapping filaments and the general complexity of the estimated structures.

\ \\ \noindent\textbf{V2: Interactive tuning of the model parameters.} Another requirement for supervised fitting is to expose the model parameters to the researcher. By tuning them, changes to the estimated solution can be made interactively. In conjunction with \textbf{V1}, this makes it possible to carry out a single fitting session within minutes. Providing an integrated feedback loop enables the researcher to quickly iterate on the solution, and accelerates the subsequent analysis and validation. Naturally, the employed reconstruction model has to be responsive enough to support this.

\ \\ \noindent\textbf{V3: Provide multiple rendering modalities.} Both point-based rendering and volume rendering modalities are necessary to analyze cosmological datasets and to support filament reconstruction. Point-based rendering is used to visualize the raw input data (a set of points in 3D space) as well as the discrete agents of the MCPM reconstruction algorithm (described in Sec.~\ref{Sec:SimulationAndModelFitting}). Volume rendering supports a range of different views, including: 
\begin{itemize}[leftmargin=*,nosep]
 \setlength\itemsep{0em}
    \item the 3D volumetric footprint of the input data, essentially a kernel point-density estimate, or \textit{deposit}, used to show the overall impact of the inputs, including the impact of their weights (i.e., masses);
    \item 
    a 3D density of the reconstructed estimate of the Cosmic Web itself, or \textit{trace}, used to examine the geometric structures and density distribution of the solution without the visual bias of the input data;
    \item 
    a combined 3D view of the \textit{deposit+trace}, used to visually examine the alignment between them and to check for any data not covered by the estimate.
\end{itemize}

\ \\ \noindent\textbf{V4: Statistical summarization of the model fit.} Alongside the visual analysis of the Cosmic Web reconstruction, researchers find it useful to view a concise set of statistics summarizing the solution. The quantities that have proven to be most valuable when determining the quality of the solution are:
the overall energy (fitness) of the reconstruction and its temporal evolution during a fitting session;
a density histogram across the entire 3D domain, to indicate any suspicious discontinuities or inexplicable increases in the number of modes in the distribution;
a histogram of densities specifically at the input data locations, to understand whether the estimated density field is proportional to the input data weights;
and the number of null-density points, i.e., the input data points missed by the reconstruction (ideally, this will approach zero, as the fit should cover all the data).

\ \\ \noindent\textbf{V5: Interactive highlighting of transition regions.} Of particular interest to astrophysicists is understanding the transition regions of the Cosmic Web, the subset of 3D space on the boundary of the circumgalactic medium (CGM) surrounding galaxies and the intergalactic medium (IGM) between galaxies. While models exist for describing the CGM and for predicting the statistical properties of the IGM~\cite{tumlinson2017circumgalactic}, the transition region has so far eluded deeper understanding. Visualization can facilitate the investigation of these regions, such as through using density-dependent color coding and interactive thresholding. 

\ \\ \noindent\textbf{V6: Inclusion of proxy objects.} Assisting with the fitting is the ability to use visual aids, such as sightlines or distance markers. These proxy objects need to be consistently superimposed against the volumetric data to help navigation in the 3D space, a task that can be challenging due to the complex structures obtained during the reconstruction.

\ \\ \noindent\textbf{V7: Exporting the reconstructed solution.} Once satisfied with the estimated Cosmic Web structure, researchers need the output data available in a format that supports further investigation using astrophysics analysis packages, such as \textit{yt}~\cite{turk2010yt}, \textit{pyigm}~\cite{pyigm}, and \textit{linetools}~\cite{linetools}. The output data includes a 3D table of scalar density values packed in a binary file, and associated metadata describing the parametrization of the reconstruction model and the spatial orientation of the estimate.

\begin{figure*}[tb]
 \centering 
 \includegraphics[width=\textwidth]{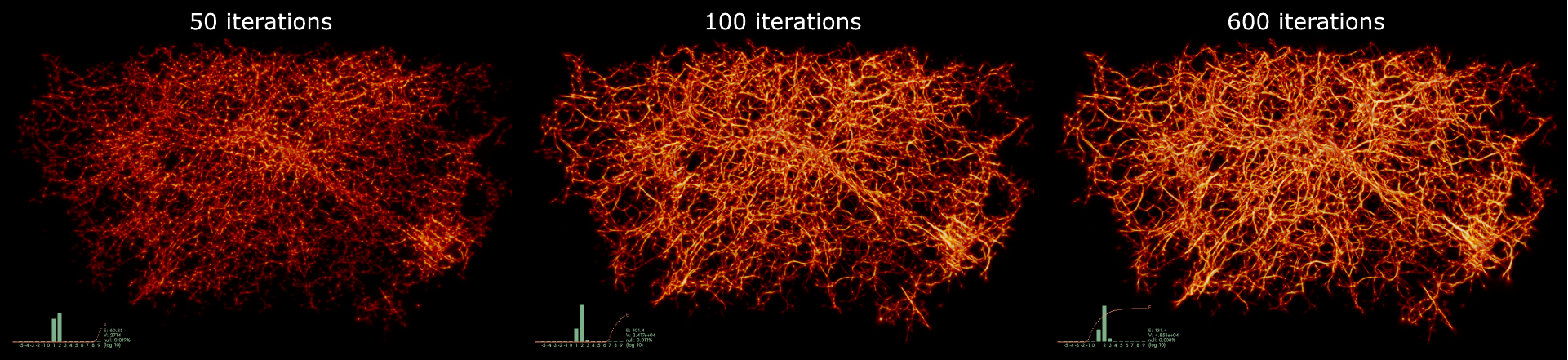}
 \vspace{-6mm}
 \caption{Progressive fitting to the SDSS galaxy data. Due to the convergence rate of MCPM, we already obtain a reasonable approximation to the final equilibrium fit in only a fraction of the total iterations, thus increasing the interactivity of the supervised fitting process.}
 \label{fig:fitting_session}
 \vspace{-2mm}
\end{figure*}

\section{Simulation and Model Fitting}
\label{Sec:SimulationAndModelFitting}

The Monte Carlo Physarum Machine (MCPM) model enables the structural analysis tasks defined above. It serves as an effective structural interpolator for cosmological datasets, which on the scales where the Cosmic Web exhibits distinct features (between 1 and 100 Mpc) can be represented as a weighted set of points in 3D space. MCPM is a dynamic model defined by a set of priors: a set of probability density functions and their parameters obtained from domain-specific knowledge and fitting to training data. Once configured, the model is fitted to the input data, resulting in a continuous geometric structure which we interpret as a transport network connecting the data points. Rather than a graph, this geometry is represented as a 3D density field stored as a regular lattice. An overview of MCPM outputs and related visualization modalities is provided in Fig.~\ref{fig:teaser}.

MCPM has both a discrete and a continuous component. The discrete component is an ensemble of particle-like agents that can freely navigate the simulation domain, representing different nuclei of the virtual organism. The continuous component is a 3D scalar lattice representing a concentration of a marker (also referred to as a ``chemo\-attractant'' in the biological context) that facilitates communication between the agents and the data. The model's behavior is based on a feedback loop cycling between these two components, executed in two alternating steps, obtaining approximations from an iterated composition of simple, local rules (which are illustrated in Fig.~\ref{fig:prop_steps}).

\ \\ \noindent\textbf{The propagation step} is executed in parallel for each of the agents, which are the model's device for exploring the simulation domain. Each agent's state is represented by a position and a movement direction, which are stochastically updated (Fig.~\ref{fig:prop_steps}a-c) to navigate through the \textit{deposit field} (referred to as the ``trail'' in~\cite{Jones2010}, see Fig.~\ref{fig:prop_steps}d, gray). The deposit field is stored as a 3D lattice of scalar values, representing the marker emitted by both the input data points as well as the agents. The deposit guides the agents towards the input data points, as the agents move with higher likelihood to the regions with greater deposit values. In addition to the deposit, we also maintain a scalar \textit{trace field} (the green cells in Fig.~\ref{fig:prop_steps}d) which records the agents' equilibrium spatio-temporal density, but which does not participate in their decision making about which direction to move in. Successive propagation steps define the agents’ trajectories, which follow the structure of the deposit field, and are recorded in the trace field. While each individual agent draws a seemingly chaotic path (Fig.~\ref{fig:agent_trajectories}, left), a superposition of many agents averaged over a sufficient time window results in a smooth and converged structure (Fig.~\ref{fig:agent_trajectories}, right). 

The input data points are also represented by agents, but are of a special type which fixes their position in space and enables them continually emit an amount of deposit proportional to their weight. We found this to be the most consistent way of handling the inputs, as the relative structural impact of the agents and the data simply becomes a question of tuning the amount of deposit emitted by each type.

The agent propagation step is further defined in terms of three probability distributions, the configuration of which can be interactively tuned in runtime. These are:
\begin{itemize}[leftmargin=*,nosep]
\setlength\itemsep{0.2em}
    \item $P_{dir}$ for sampling the agents' directional decisions during the sensing and movement phases, defined on the unit sphere relatively to the current direction of propagation. 
    \item $P_{dist}$ for sampling the agents' distance decisions during the sensing and movement phases, defined in positive real numbers along the current propagating direction.
    \item $P_{branch}$ for making the binary decision whether the agent should take a newly sampled direction or remain on its current course.
\end{itemize}

\begin{figure}[tb]
 \centering 
 \includegraphics[width=\columnwidth]{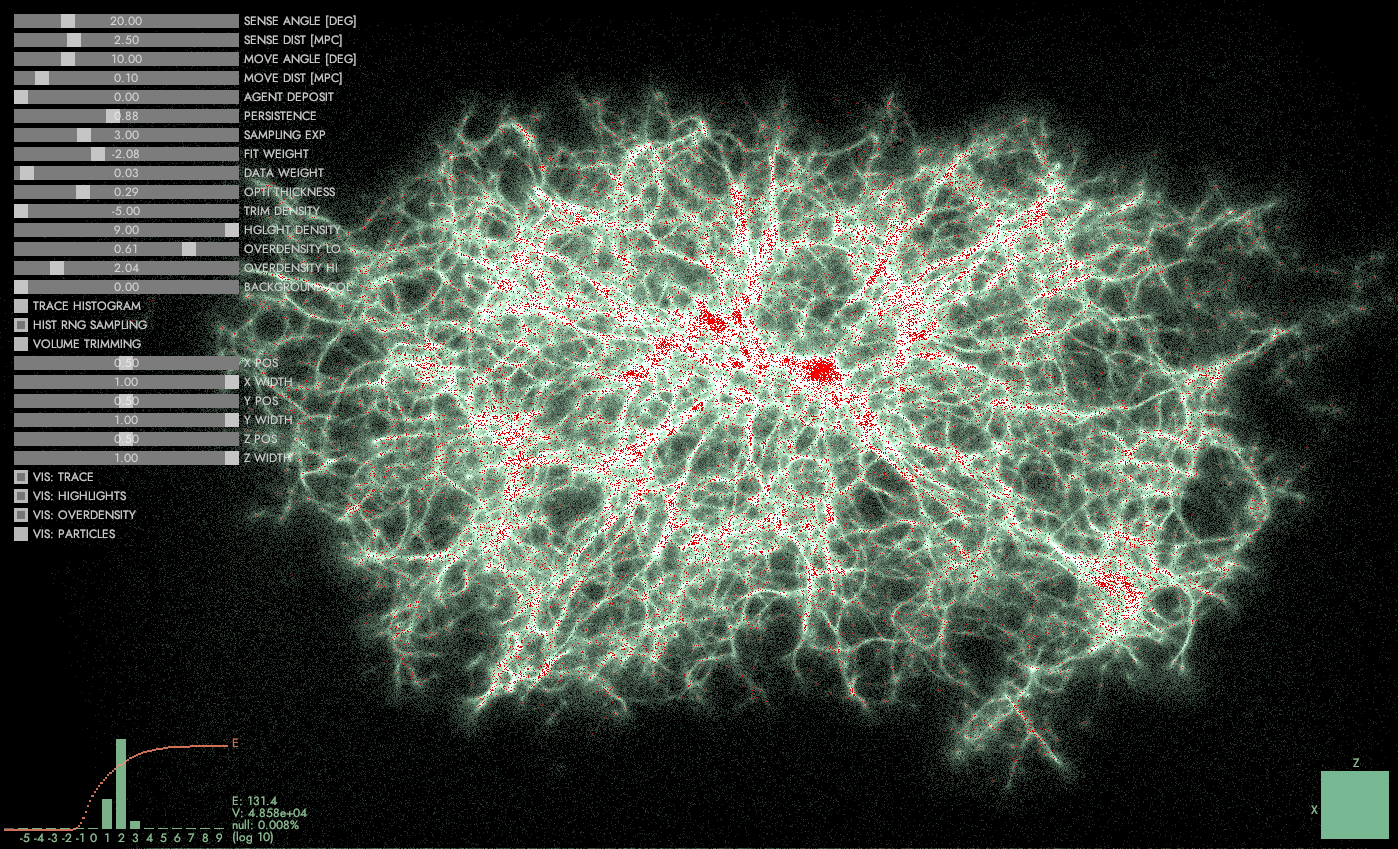}
 \vspace{-6mm}
 \caption{The initial view of \textit{Polyphorm} using our example SDSS dataset. Input data are depicted as red points and the MCPM agents as white points. The fitting starts automatically,  interactively responding to the current parametrization, which is controlled via the sliders on the top-left of the screen. The display in the bottom-left corner provides a statistical summary of the fitting, and the display in the bottom-right corner shows the current trimming state. (See Sec.~\ref{Sec:VisualAnalysis} for details.)}
 \label{fig:main_interface}
 \vspace{-6mm}
\end{figure}

\ \\ \noindent\textbf{The relaxation step} ensures that the simulation eventually reaches an equilibrium. The deposit field is spatially diffused by a small isotropic kernel, and then attenuated (i.e., each cell in the deposit field is multiplied by a value $<$ 1). The trace field is also attenuated but does not diffuse in order to preserve geometric features in the agents' distribution, which the trace ultimately represents. The simulation reaches its equilibrium when the amount of deposit and trace injected into the respective fields in the propagation step equals the amount removed by the attenuation in the relaxation step, typically within 100s of iterations.

\ \\ 
\indent In order to apply a Physarum model to the analysis of cosmological data, MCPM extends the pioneering work of Jones~\cite{Jones2010} in several ways. First, in MCPM, all the decisions made in the agent propagation step are stochastic. This makes the model highly customizable, as prior domain-specific knowledge can be encoded by probability density functions. Second, we extend Jones's model from 2D to 3D, taking advantage of the probabilistic form, and enabling analysis of higher-dimensional datasets. Third, we add a new data modality--- the trace field--- that represents the equilibrated agent density, decoupled from the algorithm’s signaling mechanism.

In Jones' model, agents always respond to different deposit concentrations by following the direction of maximum deposit concentration. For our applications, where the main concern is recovering a network structure that fits target data, we require a model that produces more complex behavior and that can be configured based on available knowledge about the data distribution. This behavior is represented by the branching decision encoded by the discrete probability $P_{branch}$, which determines whether the agent remains on its current course or instead branches out in a new direction (generated in the propagation step). To provide additional flexibility in representing paths with different scales and curvatures, we also modified the agents’ spatial decisions to behave probabilistically, rather than using constant step sizes and fixed angles for the agents’ movement. This behavior is defined by the continuous probability density functions $P_{dir}$ (2D distribution on a sphere) and $P_{dist}$ (1D distribution on a half-line). 

We also adapt and optimize Jones’ model for 3D data. Enabled by the notion of branching probability $P_{branch}$, we can now simplify the sampling stencil, that is, how many directions need to be sampled in the agent's sensing stage during the propagation step. In Jones' work, the network complexity (connectivity) depends on the number of directional samples, which becomes more pronounced in the topologically richer 3D space. Given that the directional navigation in MCPM is guided by $P_{dir}$ and $P_{branch}$, we can reduce the necessary number of samples to only \emph{two}: one in the forward direction, and one in the branching direction. This reduction in directional sampling frees up computational processing that can instead be invested into increasing the resolution of the agent trajectory, and extensions to higher dimensions are now possible without increasing the sampling load.

Further, in addition to the deposit field, MCPM defines an additional data modality: the trace field. As with the deposit, the trace is stored in a 3D lattice with the same resolution. The trace is computed as a superposition of all locations visited by each agent within a given time window, that is, it is an equilibrium distribution of agent trajectories (Fig.~\ref{fig:agent_trajectories}). Loosely speaking, the trace field can be interpreted as the total probability of the agents' distribution marginalized over all possible geometric configurations that can occur in a given dataset. The trace field has advantages over the deposit field. It is not subject to spatial diffusion and structural details are therefore preserved.

\begin{table}[b]
\small
\centering
\vspace{-4mm}
\caption{Default parameter values used in the MCPM fitting.}
\vspace{-1mm}
\begin{tabular}{r|lcr|l}
    \verb|sense_angle| & $20~[\mathrm{deg}]$ & \quad &
    \verb|sharpness| & $3$--$5$ \\
    \verb|sense_distance| & $2.5~[\mathrm{Mpc}]$ & \quad &
    \verb|agent_deposit| & $0.005$ \\
    \verb|move_angle| & $10~[\mathrm{deg}]$ & \quad &
    \verb|data_deposit| & $M_\odot\ / 10^{11}$ \\
    \verb|move_distance| & $0.1~[\mathrm{Mpc}]$ 
\end{tabular}
\label{tab:MCPM_params}
\vspace{0mm}
\end{table}

\begin{figure*}[tb]
 \centering 
 \includegraphics[width=\textwidth]{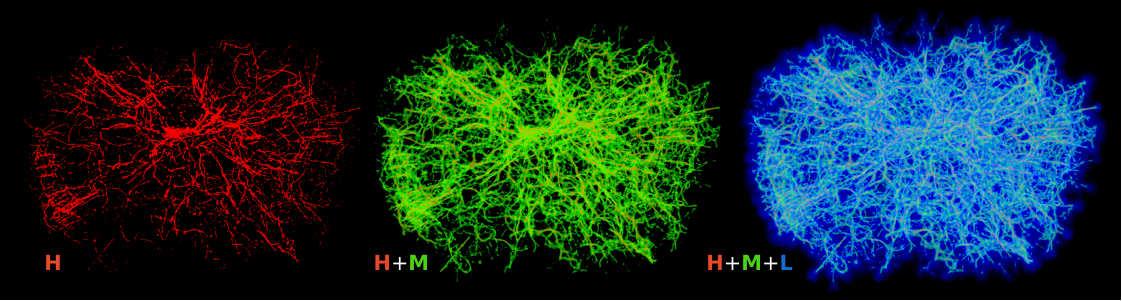}
 \vspace{-6mm}
 \caption{Segmentation of the trace into three regimes using predetermined thresholds: high density (H) corresponding to the galactic environments and circumgalactic medium, medium density (M) covering the vast majority of filaments, and low density (L) outlining outer regions of the IGM and the outskirts of voids.}
 \label{fig:overdensity}
 \vspace{-4mm}
\end{figure*}

\begin{figure}[tb]
 \centering
 \vspace{-0mm}
 \includegraphics[width=\columnwidth]{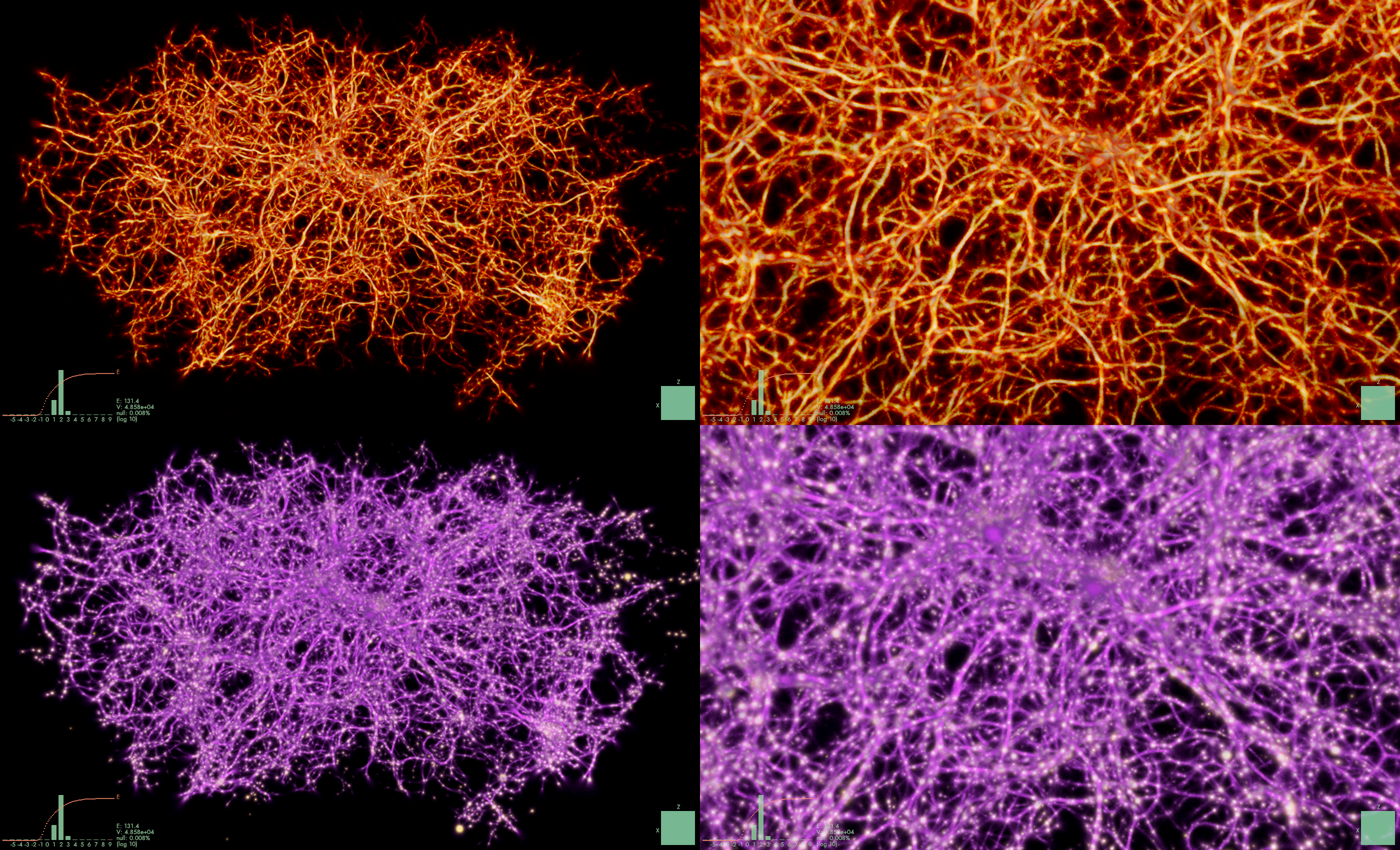}
 \vspace{-6mm}
 \caption{Top row: Converged cosmic web estimate reconstructed by \textit{Polyphorm}, mapping the underlying trace densities to a user-defined color palette for visualization. Bottom row: Converged cosmic web estimate (purple) superimposed over the data footprint (pale white) represented by the deposit field. We see a tight alignment between these structures.}
 \label{fig:trace_and_deposit}
 \vspace{-4mm}
\end{figure}

Finally, MCPM fully decouples the representation of agents from the underlying discrete data structures. In Jones’ model, only one agent could be present in each cell of the deposit field at any given time. This is suitable for simulating the actual organism, as each agent represents a small fragment of the body occupying a certain volume of space. However, for our purposes this behavior is no longer desired. Our agents represent an abstract spatial distribution of the intergalactic medium with possibly several orders of magnitude in dynamic range, and we allow agents to move freely without enforcing any limit on their number within each grid cell. 

\ \\ \indent In summary, MCPM solves tasks \textbf{S1}--\textbf{S4} by functioning as a structural interpolator for input data. When configured and parametrized appropriately (typical settings are listed in Table~\ref{tab:MCPM_params}), it reliably converges to transport networks that effectively connect neighboring data points in space. The key is that the agents are drawn towards large `pools' of deposit when $P_{branch}$ is defined such that it preferentially selects those directions where more deposit is detected. Agents are therefore attracted towards data in their vicinity (thus interconnecting them), as well as other nearby agents (thus reinforcing existing pathways). The shape characteristics of the network (such as the curvature of the paths and the size of the features, as well as connectivity patterns) are further dependent on the combined influence of $P_{dir}$ and $P_{dist}$. These networks are represented by a converged trace, that is, by a continuous density field rather than by an explicit graph structure.

\section{Visual Analysis}
\label{Sec:VisualAnalysis}

\textit{Polyphorm} is comprised of two integrated components: the simulation engine and the visualization interface. A researcher initializes the fitting with initial structural parameters (determined, e.g., from prior fitting runs), and then observes the progress of the fitting. At any point in time, based on the current simulation state, any structural or visual parameter can be adjusted until a satisfactory result is produced, at which point the researcher can continue to use \textit{Polyphorm} to investigate the reconstruction, or can export the reconstruction for further analysis.

An example of a fitting session is depicted in Fig.~\ref{fig:fitting_session}, which shows the reconstruction of the Cosmic Web from 37.6k observed galaxies from the SDSS corpus. Although a full estimate takes about 600 MCPM iterations to produce (just over 1 minute on our development machine), a significant part of the reconstruction is revealed in a fraction of that time (at around 100 iterations), allowing for quick readjustments of the model parameters without having to wait for a fully converged estimate. The interactivity of both the MCPM and the visualization enables a responsive feedback loop, reducing the time it takes to obtain a satisfactory solution from hours--- typical for other Cosmic Web reconstruction methods~\cite{Libeskind2018}--- to minutes. This provides a critical advantage when tackling a problem like this, where the solution is under-determined from the data and human supervision is required. Below, using the 37.6k galaxies extracted from the SDSS corpus, we describe the main components of the \textit{Polyphorm} interface that contribute to the reconstruction process, where each component is aligned with one or more of the the visual analysis tasks described in Sec.~\ref{Sec:VisualAnalysisTasks}. 

Upon startup, \textit{Polyphorm} displays the particle rendering mode (\textbf{V3}). The input dataset is by default visualized as red particles, while the MCPM simulation agents are white (users can customize the color mapping as desired). In our SDSS example, the simulation agents outnumber the data about 300-fold (see Fig.~\ref{fig:main_interface}). The structural and visual parameters are exposed in the top-left part of the viewport, and each of them can be adjusted interactively (\textbf{V2}). A visualization option available in all the rendering modalities that allows the researcher to slice the dataset along all 3 axes (\textbf{V1}), and in the bottom-right corner, a display indicates the current trimming state. A statistical summary of the fit (\textbf{V4}) is provided in the bottom-left corner, showing a per-galaxy trace histogram, the energy plot of the fit plus its evolution during fitting, and the number of null-density data points. In Fig.~\ref{fig:main_interface}, we can see from the energy plot that the simulation has already converged to an equilibrium, with the MCPM agents tracing the Cosmic Web estimate. At this point the simulation can be exported as a 3D density field for further analysis (\textbf{V7}).

\begin{figure}[b]
 \centering
 \vspace{-3mm}
 \includegraphics[width=\columnwidth]{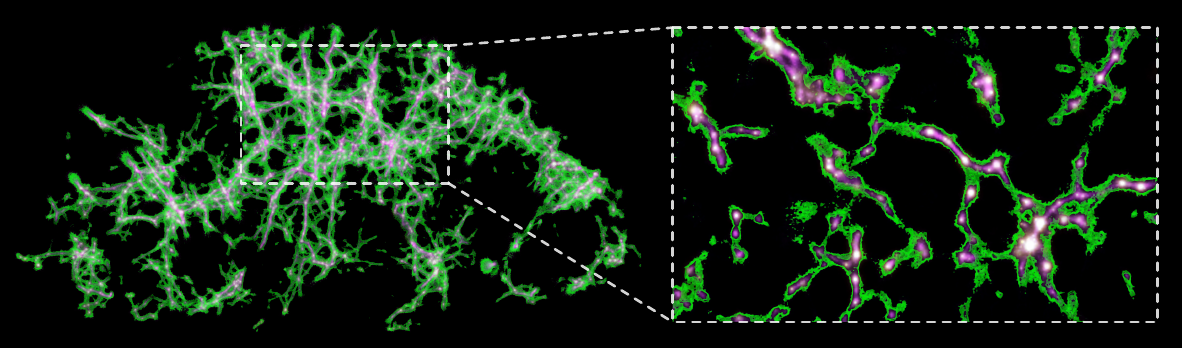}
 \vspace{-6mm}
 \caption{An example of interactive highlighting. We emphasize a particular narrow trace density range to identify the transition from the outskirts of galaxies, through the circumgalactic medium, to outer regions of space.}
 \label{fig:highlights}
\end{figure}

\begin{figure*}[tb]
 \centering 
 \includegraphics[width=\textwidth]{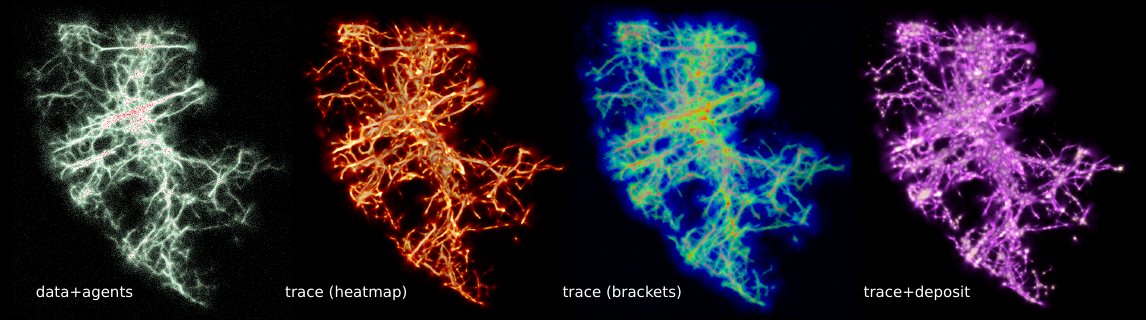}
 \vspace{-6mm}
 \caption{This figure shows an example of axis-aligned trimming. Here, we isolate a thin vertical slice to enable a closer inspection of the reconstructed structures and the trimming is applied to each of \textit{Polyphorms}'s rendering modalities.}
 \label{fig:trimming}
 \vspace{-4mm}
\end{figure*}

\subsection{Trace and Deposit Rendering Modes}

To visualize the resulting fits (\textbf{V1}), \textit{Polyphorm} uses two main views (\textbf{V3}): trace-only (Fig.~\ref{fig:trace_and_deposit}, top) and combined deposit+trace (Fig.~\ref{fig:trace_and_deposit}, bottom). Recall that the deposit represents the volumetric footprint of the input data and that the trace is the Cosmic Web density field estimated by the MCPM simulation. Both of these views are rendered via a standard volume rendering technique, using the emission+absorption model implemented through back-to-front compositing~\cite{Beyer2015}.

The trace is rendered by mapping the density field values to a customizable color map (akin to the color maps used in Matplotlib and similar packages), which therefore serves as a transfer function. In the trace-only view, we use a simple color gradient, by default starting with red at low trace values and ending in white for high values. The transparency of the medium is likewise scaled by the trace values, so that empty spaces (`voids') do not occlude the filament structure. The purpose of this view is to examine the geometry of the Cosmic Web estimate without being biased by the input data distribution.

The combined deposit+trace view is rendered using the same methodology as the trace-only view. In this view, however, the deposit (white halos) is superimposed with the trace (purple filaments). The purpose of this view is to visually verify that the trace is distributed evenly and that it aligns well with the input data.

\subsection{Trace Highlighting and Segmentation}

Complementing the global visual inspection of the trace, \textit{Polyphorm} implements additional visualization modalities: density segmentation and highlighting (\textbf{V3}, \textbf{V5}). The main purpose of the segmentation mode is to split the reconstructed trace distribution into three mutually exclusive density intervals, as shown in Fig.~\ref{fig:overdensity}. Here, the threshold values for the segmentation have been determined by our astrophysicist collaborators, relying on the calibrated mapping between the trace values and the cosmic overdensity, a well understood quantity in cosmology (see Sec.~\ref{Sec:UseCases}).

To highlight a particular narrow density interval in the trace distribution, \textit{Polyphorm} also implements thresholding in the deposit+trace modality. In Fig.~\ref{fig:highlights}, we use a trace density value that is known to correspond to the medium density on the outskirts of galactic halos and denser filaments. This highlighting can be used to verify that this spatial assumption is found in our reconstruction of the Cosmic Web.

\subsection{Axis-aligned Trimming}
\label{Sec:Trimming}

An important interactive component of \textit{Polyphorm} is the axis-aligned trimming (or `slicing') available across all of the visualization modalities (\textbf{V1}, \textbf{V3}). Having the ability to slice the data in different directions is invaluable, since spatial orientation in the full 3D dataset can be quite a challenge. Fig.~\ref{fig:trimming} demonstrates this functionality by isolating a thin vertical slice from the SDSS dataset (about 10\% of the total width). Regardless of the employed rendering modality, we can observe structures that may be occluded when viewing the reconstruction in its entirety.

\subsection{Proxy Visuals}

\textit{Polyphorm} also includes custom visualization aids (\textbf{V6}), and is designed so that additional proxy objects can be added as needed for particular datasets. These help to orient the user when navigating large 3D dataset and to better understand or to emphasize spatial relationships in the data. For example, we can add spherical shells to signify distances from Earth measured in the Cosmic redshift values. These translucent proxy objects make it apparent that the data is defined in spherical coordinates, and simplify the comparison of redshift distances of various parts in the reconstruction. Additionally, we can add one or more sightlines connecting Earth to distant objects, such as quasars or fast radio bursts (FRBs). In Use Case 3 (Sec.~\ref{Sec:UseCase3}), astrophysicists use \textit{Polyphorm} to measure the amount of dispersion in the signal arriving from a particular FRB. According to their hypothesis, this dispersion is caused by interactions with the Cosmic Web. By visualizing the FRB sightline (see Fig.~\ref{fig:frb}), astrophysicists can more easily reason about its intersections with the Cosmic Web.

\begin{table}[thb]
\small
\centering
\caption{Performance of \textit{Polyphorm} in processing datasets of different sizes (corresponding to the three use cases detailed in \protect{Sec.~\ref{Sec:UseCases}}). All three experiments use 10M MCPM agents and a screen resolution of 1400$\times$900. We list: the number of input data points; total number of voxels in the deposit and trace grids and their actual resolutions; total amount of video memory consumed; MCPM simulation time per iteration; rendering time per frame; total time and number of MCPM iterations until convergence. The rendering times vary within the intervals based on the chosen modality, visualization settings, and camera position.}
\label{tab:performance}
\begin{tabular}{r@{\hspace{0.2cm}} ccc}
     & Small (UC3, FRB) & Medium (UC2, SDSS) & Large (UC1, BP) \\
    \toprule
    Data points: & 1.3k & 37.6k & 840k \\ \hline
    \multirow{2}{*}{Voxels:} & 324M & 670M & 2148M \\ 
    \quad & (2$\times$1024$\times$600$\times$264) & (2$\times$560$\times$1024$\times$548) & (2$\times 1024^3$) \\  \hline    Memory: & 1GB & 2GB & 6GB \\ \hline
    Simulation: & 55\,ms & 80\,ms & 165\,ms \\ \hline
    Rendering: & 15--45\,ms & 20--160\,ms & 30--220\,ms \\ \hline
    \multirow{2}{*}{Convergence:} & 15\,s & 75\,s & 145\,s \\ 
    \quad & (200 iterations) & (600 iterations) & (650 iterations) \\
    \bottomrule
\end{tabular}
\vspace{-5mm}
\end{table}

\subsection{Implementation Details}

\textit{Polyphorm} was written in C++ (for the CPU code) and DirectX's HLSL (for the GPU code). All performance-critical steps are parallelized and implemented as HLSL compute shaders, including MCPM's propagation and relaxation steps, as well as the particle and volume rendering modes. All data structures in \textit{Polyphorm} are ordered, with emphasis on fast access in both sequential and random modes. Specifically, the input data points (galaxies or dark matter halos) and MCPM agents are stored in linear arrays, using \texttt{fp32} to store their positions, weights, and, in the case of agents, directions. The deposit and trace fields are stored in regular 3D grids with \texttt{fp16} voxels for efficiency reasons (as they represent the primary memory bottleneck).

Table~\ref{tab:performance} lists performance figures for \textit{Polyphorm} using the three datasets analyzed in the Scientific Use Cases below: Bolshoi-Planck (large dataset, Sec.~\ref{Sec:UseCase1}), SDSS (medium dataset, Sec.~\ref{Sec:UseCase2}) and FRB (small dataset, Sec.~\ref{Sec:UseCase3}). All results were obtained using a desktop computer outfitted with an Intel Core i9-9900K CPU and an NVIDIA TitanX GPU. The simulation and visualization performance scales approximately linearly with the number of voxels, as expected, and sub-linearly with the number of agents (e.g., using 100M agents instead of 10M incurs only a 2x slowdown). Although using higher numbers of agents has only a small impact on the model's spatial details, this does improve the model's effective resolution by reducing the Monte Carlo noise levels. Using 10M agents has proven sufficient across all our use cases.

\begin{figure}[tb]
 \centering
 \includegraphics[width=\columnwidth]{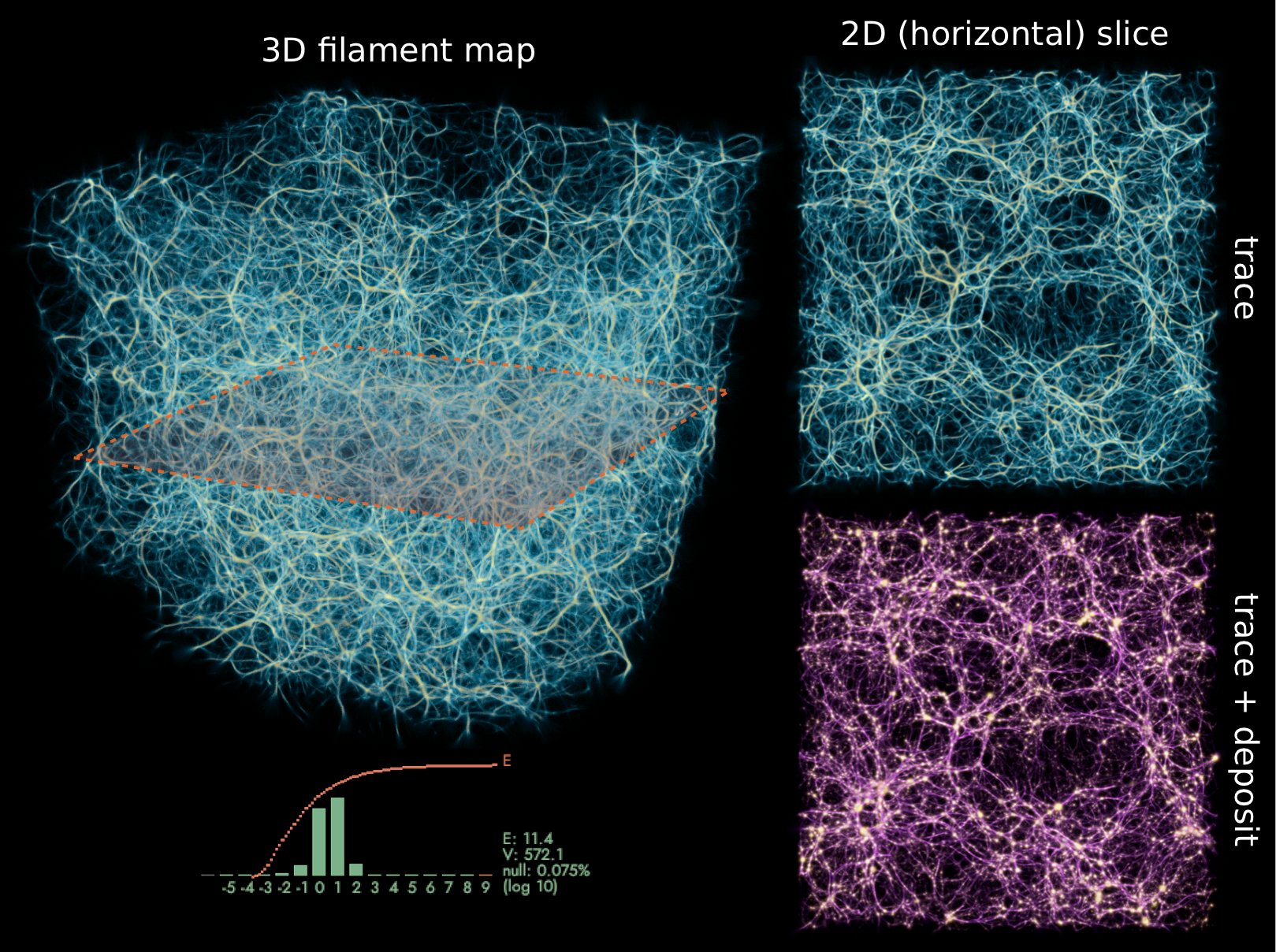}
 \vspace{-6mm}
 \caption{Reconstruction of dark matter filaments in the simulated Bolshoi-Planck dataset (containing 840k halos). \textit{Polyphorm} yields a consistent 3D structure, enabling its calibration to cosmic overdensity values (\protect{Sec.~\ref{Sec:UseCase1}}). Thin slices of the filament map are shown on the right, obtained with the help of the trimming functionality (see Sec.~\ref{Sec:Trimming}).}
 \label{fig:bolshoi_planck}
 \vspace{-4mm}
\end{figure}

\section{Scientific Use Cases} 
\label{Sec:UseCases}

\newcommand{\rhom}{\rho_m}
\newcommand{\rhomm}{\langle \rhom \rangle}
\newcommand{\rhophys}{\rho_{\mathrm{Phys}}}

\subsection{Use Case 1: Establishing a mapping between the MCPM density estimate and cosmic overdensity}
\label{Sec:UseCase1}

\textit{Polyphorm} was used to establish a complete, calibrated map of the dark matter density field (traced by the locations of galaxies) in a sufficiently representative region of space in order to study the distribution of IGM gas relative to the dark matter structures~\cite{Burchett2020}. As current consensus estimates that the sizes of IGM filaments range between a few to dozens of Mpc, our reconstruction needed to encompass a region of space larger than this. Similarly, an effective spatial resolution is needed to represent scales as small as the extent of typical galactic halos, between 0.1 and 1 Mpc.

The trace produced by the MCPM model defines a 3D density field $\rhophys$ (\textbf{S1}). To calibrate these values to the more cosmologically meaningful cosmic matter density term $\rhom$, our collaborators used \textit{Polyphorm} to fit the MCPM model to the halo catalog produced by the ROCKSTAR algorithm~\cite{Behroozi:2013aa} for the Bolshoi-Planck (BP) simulation at the $z=0$ redshift snapshot. The resulting fit is shown in Fig.~\ref{fig:teaser} in the different visual modalities provided by the \textit{Polyphorm} visualization engine (\textbf{V3}). As per requirements defined by the structural tasks \textbf{S2} and \textbf{S3}, the reconstruction yields the expected anisotropic filamentary structures across multiple scales in space and density (see Fig.~\ref{fig:bolshoi_planck}). The reconstruction also has excellent alignment with the 840k dark matter halos in the BP dataset, and only a negligible number of halos ($<0.1$\%) are not included in the reconstruction. The spatial requirements were likewise fulfilled: the reconstruction spans a cubic volume 170 Mpc in each side, with a density field resolution of $1024^3$, i.e., the minimum represented feature size was 0.166 Mpc.

The calibration to cosmic matter density $\rhom$ then involved dividing the $\rhophys$ values into a histogram with bins of width 0.05 in the log$_2$ domain, and spatially cross-matching cells in each $\rhophys$ bin with a reference $\rhom$ field, which is part of the BP corpus. This reference 3D matter density field was derived by smoothing the particle mass density in the BP volume over 0.25 Mpc scales, similar to the resolution scale of our \textit{Polyphorm} data product.  The actual mapping was then obtained via the running median method~\cite{astola1989computation} on the cross-matched densities $\rhophys$ and $\rhom$, and subsequent normalization by the mean matter density to obtain a relative quantity, cosmic overdensity $\rhomm$. It is important to note that while a more precise map could have been obtained by fitting to this reference $\rhom$ field, this was explicitly avoided since fitting to sparse point data (akin to the observational data where only galaxy locations are known) was the key task \textbf{S1} of \textit{Polyphorm}.

The MCPM trace and BP matter densities are well correlated for log$_2$ $\rhophys$ $\gtrsim 1$, which was the regime of interest for our astrophysicist collaborators. The use of \textit{Polyphorm} was a critical ingredient in establishing a first-of-its-kind, continuous spatial map between a synthetic computational model (MCPM) and a canonical cosmological quantity, the cosmic overdensity $\rhom/\rhomm$.

\subsection{Use Case 2: Reconstructing the Cosmic Web from SDSS galaxy observations for IGM analysis}
\label{Sec:UseCase2}

Based on the results of Use Case 1, along with the fact that the MCPM model yields a 3D density distribution $\rhophys$ (\textbf{S1}), it now becomes possible to analyze astrophysical observables in context with the large scale structure. The MCPM trace density field data product readily enables analyzing correlations between, e.g., galaxies' star formation rates and their local environmental density. This may be achieved in a straightforward manner because a local MCPM density value can be assigned to each galaxy based solely on its spatial coordinates. \textit{Polyphorm} was used to track the density of atomic hydrogen (H I) as a tracer of the IGM, testing the long-held assumption that the spatial distribution of H I follows that of the overall Cosmic Web structure. To map out the predicted structure, our collaborators used 37.6k galaxies from the Sloan Digital Sky Survey (SDSS) dataset, spanning across 280 Mpc of space at redshifts $z = 0.0138-0.318$. (This is also the dataset used for Figs.~5--10 referenced in Sec.~\ref{Sec:VisualAnalysis})

To obtain a robust mapping to the SDSS data (which are relatively sparse), visual analysis tasks \textbf{V1}--\textbf{V4} were employed. The structural transfer (\textbf{S4}) facilitated by the prior fitting to the BP data was key to estimating the MCPM parameters for this fitting. The primary technique for studying H I in the IGM is quasar absorption line spectroscopy, in which light from distant quasars serves as background radiation sources~\cite{richards2002spectroscopic,spitzer1959interstellar,dieter1966recent}. Elements and ions, such as H I, leave imprints on the background light in the form of spectral absorption features (or lines) at the specific locations they pass through. To detect and measure the line-of-sight density of the IGM gas, 346 spectroscopic sightlines of distant background quasars piercing the MCPM-mapped SDSS region were obtained from the Hubble Space Telescope archival database~\cite{Peeples:2017aa}. Our astrophysicist collaborators preprocessed the spectra from this database to search for the H I Ly$\alpha$ absorption signature by parametrizing each pixel by proxy for its distance from Earth along the line of sight. To avoid pollution of the absorption signal, they masked the specific wavelength ranges corresponding to the signatures of the Milky Way's inner interstellar medium. These `cleaned' spectra were then back-projected into 3D space and correlated with the Cosmic Web estimate provided by \textit{Polyphorm} (\textbf{V7}) and mapped from $\rhophys$ to $\rhom$. Binning the resulting pairs of log~$\rhom$ density + Ly$\alpha$ absorption averaged over all 346 sightlines, our collaborators found a clear relationship between these two quantities~\cite{Burchett2020}. They observed a statistically significant increase of the detected Ly$\alpha$ absorption coincident with the cosmic overdensity increasing over the mean (log~$\rhom / \rhomm > 0$). This relation demonstrates that \textit{Polyphorm} is capable of producing a viable predictor for the H I distribution in the density regime $0~\lesssim~\mathrm{log} \rhom/\rhomm~\lesssim~2$. Interactively highlighting transition regions (\textbf{V5}) provided the verification that this density regime corresponds to the circumgalactic environments as expected.

A leveling off of H I at $\mathrm{log} \rhom/\rhomm~\gtrsim~2$ and downswing observed at $\mathrm{log} \rhom/\rhomm~\gtrsim~3$ is counterintuitive, and the astrophysics community is still debating the cause of this behavior. The initial hypothesis presented by our collaborators was that, in this high-density regime, the increasing overall density is offset by the neutral H I becoming ionized due to shock heating processes and/or excess radiation from nearby galaxies. That is, the gas probed by the quasar sightlines contains fewer neutral atoms and therefore absorbs less in Ly$\alpha$ transition. \textit{Polyphorm} facilitated the visual analysis of the three main density regimes (\textbf{V3}, \textbf{V5}), including: voids, the vast under-dense pockets in the Cosmic Web where no H I signature is detected; filaments, where we observe a monotonic increase in the absorption with increasing cosmic density; and high-density shock-heated regions, where the increasing absorption signature is eventually suppressed. These results provide strong evidence towards the conjecture that the bulk of the IGM resides in the Cosmic Web filaments~\cite{Burchett2020}.

\begin{figure*}[tb]
 \centering 
 \includegraphics[width=\textwidth]{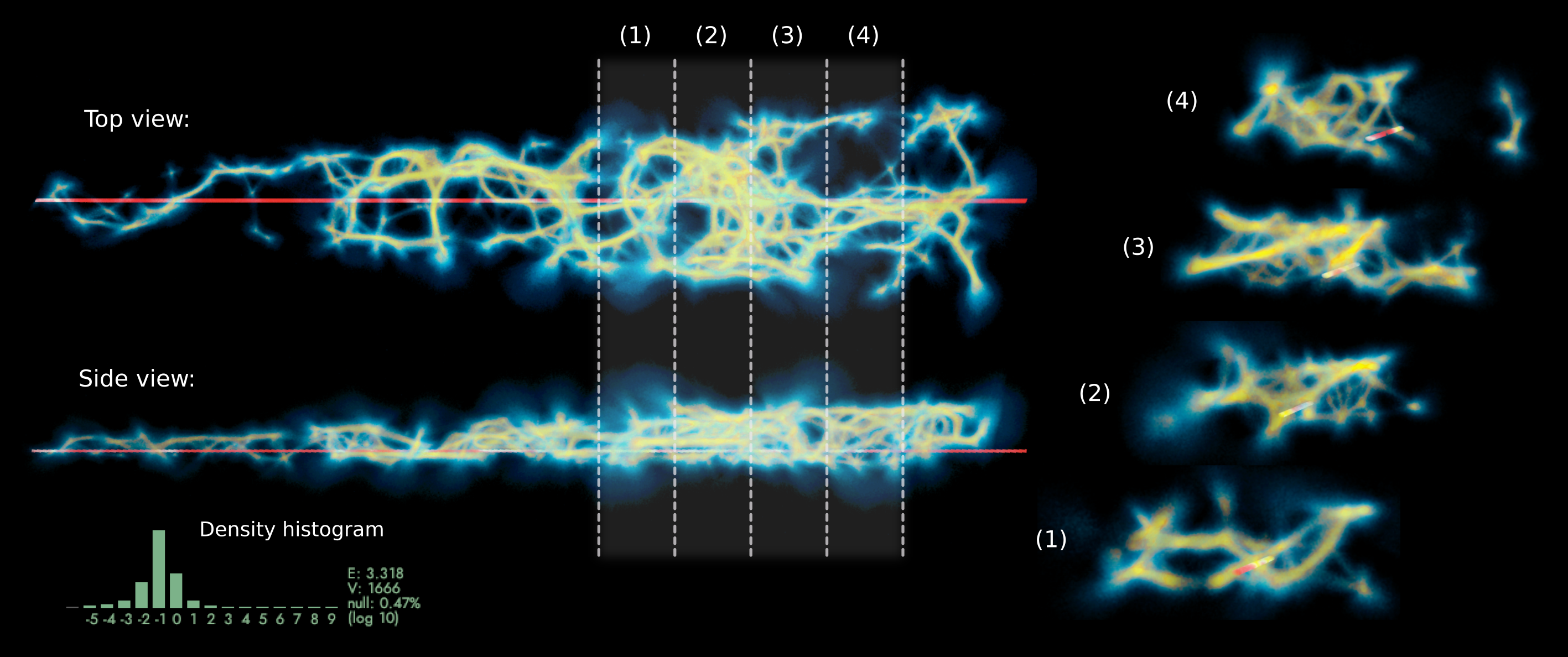}
 \vspace{-6mm}
 \caption{A 3D model of the Cosmic Web reconstructed using MCPM to support Use Case 3 (Sec.~\ref{Sec:UseCase3}). Top left: The red line represents the FRB sightline; light is assumed to travel from right to left. The Cosmic Web reconstruction is shown color-coded by the steady-state Physarum particle trace density (yellow being high and black being low). Bottom left: A rotated view of the reconstruction. The FRB sightline is along one of the narrow strips in the SDSS footprint and therefore the vertical size in the side view is smaller than that that in the top view. Right: Four cross-sectional views at the locations indicated by the dotted lines in the plots on the left. }
 \label{fig:frb}
 \vspace{-5mm}
\end{figure*}

\subsection{Use Case 3: Inferring the ionized IGM contribution to the dispersion measure of FRB 190608}
\label{Sec:UseCase3}

\textit{Polyphorm} was used to investigate the filamentary structure of the Cosmic Web in a region along a Fast Radio Burst (FRB) sightline, for which observed data indicate an unusually large dispersion measure compared to the cosmic average at its redshift, as well as an unusually large rotation measure and pulse width. By producing a 3D map of ionized gas in the filaments, our collaborators were able to compute a dispersion measure contribution from matter outside halos, leading to the first detailed end-to-end study of matter along an FRB sightline~\cite{Simha_Disentangling2020_ApJ}.

In so doing, our collaborators were able to ``disentangle'' the Cosmic Web by correlating the dispersion measure of FRBs with the distributions of foreground galaxy halos \cite{mcquinn14,xyz19}. This dispersion measure (DM) is the path integral of the electron density $n_e$ weighted by the scale factor $(1+z)^{-1}$,  ${\rm DM} \equiv \int n_e \, ds / (1+z)$. That is, these FRB measurements are sensitive to all of the ionized gas along the sightline, and therefore can be used to trace the otherwise invisible plasma surrounding and in-between galaxy halos~\cite{jpm+2020}. The DM is an additive quantity, and for an FRB it may be split into individual contributions of intervening, ionized gas reservoirs, including a contribution from the Milky Way, from the host galaxy (where the FRB originates), and from the Cosmic Web (i.e., the intervening halos made up of CGM gas found within dark matter halos adjacent to galaxies and IGM gas found between galaxy halos). Since there are no known galaxies directly intercepted by the line of sight between the Milky Way and the host galaxy, FRB 190608 lended itself to an investigation of how filaments in the Cosmic Web affect the DM. The FRB sightline falls within a region that serendipitously is covered by SDSS, providing the opportunity to consider the cosmic gas residing within the underlying large-scale structure. Theoretical models predict shock-heated gas within the Cosmic Web as a natural consequence of structure formation~\cite{Dave+2001,Cen:1999yq}, and FRBs offer one of most promising paths forward in detecting this elusive material.

The MCPM methodology produces a continuous 3D density field defined even relatively far away from galaxies on Mpc scales (\textbf{S1}), and traces anisotropic filamentary structures on both large and small scales (\textbf{S2}, \textbf{S3}). Using the SDSS galaxy distribution within 400 arcmin of the FRB sightline (1111 galaxies with mass $10^{10}$ $M_\odot$ or higher) as input to \textit{Polyphorm}, our astrophysicist collaborators used MCPM reconstruction to map the large-scale structure intercepted by the FRB sightline. The resulting 3D map enabled astrophysicists to identify and characterize the density of filaments between Earth and FRB 190608 via a similar calibration procedure as that described in Use Case 1. With this density estimate, the contributions of the ionized gas within the Cosmic Web to the observed DM could be calculated. The reconstruction of the structure intercepted by the FRB sightline is depicted in Fig.~\ref{fig:frb}. Visual analysis features that were especially important for this use case were the ability to interactively observe the reconstructed Cosmic Web in different modalities (\textbf{V3}) and to focus on particular slices of the dataset and filament reconstruction (\textbf{V1}), as well as the ability to include visual annotations (\textbf{V6}) and to export the reconstructed density field (\textbf{V7}) for subsequent analysis of the observed DM along the FRB sightline.


\section{Conclusion and Future Work}
\label{Sec:Conclusion}

In this paper, we introduced \textit{Polyphorm}, an interactive visualization tool that enables the application of the MCPM model to large cosmological datasets, and provided three scientific use cases in which the structural and and visual analysis tasks enabled by \textit{Polyphorm} led directly to new insights into composition and structure of the Cosmic Web. We will continue to use \textit{Polyphorm} to explore additional astrophysics datasets in the future, and look for alternative methods to validate the MCPM model in other cosmological contexts, using both observed and simulated data. We plan to further investigate the mathematical foundations for our stochastic model and to include additional analysis methods in future iterations of \textit{Polyphorm}, such as topological analyses for graph extraction, which is directly applicable to cataloging Cosmic Web filaments, and which may expand the range of structural tasks our software could support. Though our visualization and interaction modalities were effective in facilitating the scientific use cases we described, incorporating a richer palette of rendering techniques could provide additional insight into the input data and filament reconstructions, supporting further explorations of the transitional regions in the Cosmic Web. 

The fitting of MCPM is a dynamic process with rate of convergence following a Pareto-like distribution, which is expected from a stochastic Monte Carlo simulation. In addition, the model is continuous in space and time, including the discretization of agent trajectories (the scale of which is now related to sampling density). We believe that this makes MCPM suitable for seamless fitting of time-dependent data. If the temporal evolution of the dataset is matched with the model dynamics, the agents can be made to travel in space-time with minimal distortion, which could provide an effective way to represent the temporal evolution of the Universe. Future work will explore extending \textit{Polyphorm} to incorporate dynamic data. 

\textit{Polyphorm} is available via our open source GitHub project repository at \url{https://github.com/CreativeCodingLab/Polyphorm}, along with source code, detailed instructions on how to load in and analyze custom datasets, a video overview, and additional documentation.

\acknowledgments{
This research was supported by NASA through award HST-AR
\#15009 from the Space Telescope Science Institute. Hardware was provided by the NVIDIA GPU Grant program. Special thanks to Jan Ivaneck\'{y} for his help in developing an initial prototype of the software.}

\bibliographystyle{abbrv-doi}

\bibliography{Polyphorm}
\end{document}